\documentclass[conference]{IEEEtran}
\IEEEoverridecommandlockouts
\usepackage{cite}
\usepackage{amsmath,amssymb,amsfonts}
\usepackage{algorithmic}
\usepackage{graphicx}
\usepackage{textcomp}
\usepackage{xcolor}

 \usepackage{authblk}

\def\BibTeX{{\rm B\kern-.05em{\sc i\kern-.025em b}\kern-.08em
    T\kern-.1667em\lower.7ex\hbox{E}\kern-.125emX}}
    
\usepackage{fancyhdr}
\usepackage{kantlipsum}

\fancypagestyle{plain}{
  \fancyhf{}
  \fancyhead[C]{test}     
  \fancyfoot[L]{This is a notice}

}
\usepackage{eso-pic}

\usepackage{scalerel}
\usepackage{tikz}
\usetikzlibrary{svg.path}

\definecolor{orcidlogocol}{HTML}{A6CE39}
\tikzset{
  orcidlogo/.pic={
    \fill[orcidlogocol] svg{M256,128c0,70.7-57.3,128-128,128C57.3,256,0,198.7,0,128C0,57.3,57.3,0,128,0C198.7,0,256,57.3,256,128z};
    \fill[white] svg{M86.3,186.2H70.9V79.1h15.4v48.4V186.2z}
                 svg{M108.9,79.1h41.6c39.6,0,57,28.3,57,53.6c0,27.5-21.5,53.6-56.8,53.6h-41.8V79.1z M124.3,172.4h24.5c34.9,0,42.9-26.5,42.9-39.7c0-21.5-13.7-39.7-43.7-39.7h-23.7V172.4z}
                 svg{M88.7,56.8c0,5.5-4.5,10.1-10.1,10.1c-5.6,0-10.1-4.6-10.1-10.1c0-5.6,4.5-10.1,10.1-10.1C84.2,46.7,88.7,51.3,88.7,56.8z};
  }
}

\newcommand\orcidicon[1]{\href{https://orcid.org/#1}{\mbox{\scalerel*{
\begin{tikzpicture}[yscale=-1,transform shape]
\pic{orcidlogo};
\end{tikzpicture}
}{|}}}}

\usepackage{hyperref} 

\begin{document}

\AddToShipoutPictureBG*{%
  \AtPageUpperLeft{%
    \setlength\unitlength{1in}%
    \hspace*{\dimexpr0.5\paperwidth\relax}
    \makebox(0,-0.5)[c]{\footnotesize This is the author's version of an article that has been accepted for publication in the 2020 IEEE 29th International Symposium on Industrial Electronics (ISIE).}%
}}

\AddToShipoutPictureBG*{%
  \AtPageUpperLeft{%
    \setlength\unitlength{1in}%
    \hspace*{\dimexpr0.5\paperwidth\relax}
    \makebox(0,-0.8)[c]{\footnotesize Changes were made to this version by the publisher prior to publication. The final version of record is available at \href{https://doi.org/10.1109/ISIE45063.2020.9152371}{https://doi.org/10.1109/ISIE45063.2020.9152371}.}%
}}

\AddToShipoutPictureBG*{%
  \AtPageLowerLeft{%
    \setlength\unitlength{1in}%
    \hspace*{\dimexpr0.5\paperwidth\relax}
    \makebox(0,0.75)[c]{\footnotesize Copyright (c) 2020 IEEE. Personal use is permitted. For any other purposes, permission must be obtained from the IEEE by emailing \href{pubs-permissions@ieee.org}{pubs-permissions@ieee.org}.}%
}}

\title{Integrating 2D and 3D Digital Plant Information Towards Automatic Generation of Digital Twins}

\author[1]{\small Seppo Sierla}
\author[1]{\small Mohammad Azangoo}
\author[2]{\small Alexander Fay}
\author[1, 3]{\small Valeriy Vyatkin}
\author[4]{\small Nikolaos Papakonstantinou}

\affil[1]{\textit {\footnotesize Department of Electrical Engineering and Automation, Aalto University, Espoo, Finland}}
\affil[2]{\textit {\footnotesize Department of Automation Engineering, Helmut Schmidt University, Hamburg, Germany}}
\affil[3]{\textit {\footnotesize Department of Computer Science, Electrical and Space Engineering, Luleå University of Technology, Luleå, Sweden}}
\affil[4]{\textit {\footnotesize VTT Technical Research Centre of Finland Ltd, Espoo, Finland}}
\affil[ ]{\text {seppo.sierla@aalto.fi, mohammad.azangoo@aalto.fi, alexander.fay@hsu-hh.de, valeriy.vyatkin@aalto.fi, nikolaos.papakonstantinou@vtt.fi}}

\maketitle

\begin{abstract}
Ongoing standardization in Industry 4.0 supports tool vendor neutral representations of Piping and Instrumentation diagrams as well as 3D pipe routing. However, a complete digital plant model requires combining these two representations. 3D pipe routing information is essential for building any accurate first-principles process simulation model. Piping and instrumentation diagrams are the primary source for control loops. In order to automatically integrate these information sources to a unified digital plant model, it is necessary to develop algorithms for identifying corresponding elements such as tanks and pumps from piping and instrumentation diagrams and 3D CAD models. One approach is to raise these two information sources to a common level of abstraction and to match them at this level of abstraction. Graph matching is a potential technique for this purpose. This article focuses on automatic generation of the graphs as a prerequisite to graph matching. Algorithms for this purpose are proposed and validated with a case study. The paper concludes with a discussion of further research needed to reprocess the generated graphs in order to enable effective matching.

\end{abstract}


\section{Introduction} \label{section 1}

A major goal of Industry 4.0 is to make plant information available to humans and machines throughout the network of enterprises involved in designing, commissioning and operating the plant \cite{Zezulka}. This information includes design information as well as information gathered during the operation of the plant, so a life cycle wide information management strategy and supporting tool chains are required \cite{Harrison}. However, plant design information still often resides in proprietary and tool specific formats. A few exceptions exist, such as the Proteus XML schema for P\&ID (Piping \& Instrumentation Diagram) diagram exchange, which is supported by several leading P\&ID tool vendors \cite{Papakonstantinou}, and the PCF (Piping Component File) format for 3D isometrics, supported by leading tool vendors such as Hexagon PPM, Autodesk, Alias and PTC Creo. However, these are solutions for exchanging a specific type of diagram between tools of different vendors. There is also a need to integrate the information produced with different types of tools. The scope of this article is 2D information from P\&IDs and 3D information from CADs, in the Proteus XML and PCF formats, respectively. Our use cases for integrating such information is the generation of a digital twin, extending recent work \cite{Martinez2} by combining the control loop information from the P\&ID with the physical layout from the 3D CAD.
A straightforward approach for information integration would be to match tag names from the 2D and 3D information sources to identify the parts of these models that correspond to the same component. However, with industrial design repositories it cannot be assumed that consistent naming conventions have been enforced to enable this approach \cite{Rantala}. Thus, the integration of 2D and 3D plant information is a challenging task, so it is helpful to break it down to a process consisting of several steps. Further research on proposing the steps of such a process is solicited from other research groups. In this paper, the following process is suggested:
\begin{enumerate}
\item Digitize the information to a standard, industrially accepted Industry 4.0 format. This may involve no effort if the designs were made in tools that support these formats. However, industrial plants have lifecycles of several decades, in which case innovative applications are required to digitalize the legacy design information. Such work has been done for P\&IDs \cite{Barth,Arroyo,Nurminen}.
\item Raise the level of abstraction of the 2D and 3D designs, so that they are at the same level of abstraction.
\item Match the models generated in step 2, to identify the elements in these models that correspond to the same plant component, such as a tank or pump.
\item Use the matches to augment applications relying only on either 2D or 3D information sources. For example, \cite{Martinez3} generate the physical aspect of a digital twin of a process based solely on the 3D information, so control software is not generated and a legacy control system is expected to be integrated as in \cite{Martinez1}. \cite{Papakonstantinou} generate the cyber aspect of a digital twin, i.e. a control system, based on the 2D information. If the instrumentation in the 2D and 3D sources could be matched, it would be possible to automatically identify and connect the I/O interface of the physical and virtual aspects of the digital twin, eventually aiming at automatic generation of a system that could be considered a fully-fledged digital twin \cite{Koulamas}.
\end{enumerate}

In this paper, it is expected that step 1 has been performed. The research goal of this paper is step 2. Steps 3 and 4 are presented for motivational purposes and they are left for further research.

\section{Related work}

Significant prior work has been done with respect to step 1 of the process proposed in section \ref{section 1}. Legacy engineering documents can be digitized by scanning and storing \cite{Radial}. OCR (Optical Character Recognition) can be used to identify e.g. tag names in scans and, thus, to link engineering documents (such as equipment data sheets, work instructions, and operating manuals) that are related to each other. However, graphical data cannot be transformed into information, and links between items on a drawing cannot be transformed into a digital structural model. Several authors have investigated the extraction of text annotations from mixed text-graphic documents. \cite{Chowdhury} proposed a method for string separation in images with annotations. \cite{Gao} introduced a raster-based method for the identification of string boxes. \cite{Chai} proposed a hybrid algorithm for the same task. \cite{Wenyin} presented a method for the recognition of both text and basic parametrical forms in documents. \cite{Lu,Han} addressed the recognition of text from drawings. In \cite{carlos1}, authors presented approaches for detection and segmentation of complex engineering drawings consisting of textual and graphical elements, aiming at identification of key elements only. Also, they published a comprehensive survey on alternative approaches for the digitisation of complex engineering drawings \cite{carlos2}. Other works have focused on the analysis of symbols (OSR), which is relevant e.g. in mechanical engineering to interpret and convert design drawings \cite{Henderson}. The ultimate goal is to generate 2D or 3D models in a neutral format. \cite{Yu} presented a system which is able to interpret a range of engineering documents, such as logical diagrams, electrical circuits, and P\&IDs. This approach does not support key geometric features such as scaling, rotation, and partial overlap of objects. \cite{Berbar} presented a method to analyse design drawings, esp. electric wiring diagrams. \cite{Soon} proposed the combination of geometric and semantic information for the reconstruction of 3D CAD models from engineering drawings. The semantic information used in this approach is, however, limited to the recognition of symbols and does not consider semantic properties of the analyzed structural items. In addition, commercial methods exist which allow for automatic conversion of CAD designs into object-oriented models \cite{next}, but this requires access to the original CAD model software and can therefore not be applied to the typical use case where a plant owner has to rely on PDF documents. In\cite{Arroyo}, a method is described which combines OSR with semantic knowledge. This method allows extracting a structural model from a given 2D diagram, e.g. a P\&I diagram or a control logic diagram. Furthermore, a method is described which merges the 2D P\&I Diagram and the 2D control logic diagram into a single structural model. The method has been applied successfully to interpret engineering documents from an oil rig in the North Sea \cite{Hoernicke}. The method is limited as it relies on a consistent, common naming scheme of the tag names in both diagrams.  In \cite{Tan}, a P\&I diagram is analysed for design faults based on the identified objects and their connections. A similar approach has been patented recently by T. Tung \cite{Tung}.

For step 2 of the process proposed in section \ref{section 1}, several authors have identified graph formats as a suitable abstraction of complex engineering drawings. In \cite{Schmidberger}, it has been described how to formulate rules which can be applied to convert structural plant models into more abstract models. For example, a P\&ID which contains tanks, nozzles, pipes and joints can be converted into a structural model which provides all possible flow paths between a given set of tanks. \cite{Beez} presents an application for the automatic generation of bond graph models from an IEC 62424 hierarchical representation of the process plants. Also, \cite{Rantala} convert 3D pulp\&paper plant designs to graphs in order to perform graph matching to identify similar, and thus reusable designs. \cite{Rahul} has presented an approach for extracting information from P\&ID sheets by using deep learning networks and low-level image processing techniques for capturing inlets, outlets and pipelines as a tree-like data structure. \cite{son} uses graph abstractions to identify differences between process designs, as captured in 3D CAD models, and the as-built version of the plant, as captured by laser scans. 

In recent years, many efforts have been made to standardize process presentation formats. The ISO 15926 standard information model with its Proteus XML file format \cite{DEXPI,Fiatech} focuses on the interoperability of P\&IDs. A working group of owner operators, software vendors and research organizations called DEXPI developed a specification (DEXPI) based on the ISO 15926 to address practical issues and push for the adoption of the DEXPI/ISO15926 as an open P\&ID storage format. DEXPI and OPC Foundation have formed a joint working group for defining a DEXPI OPC UA companion specification \cite{Wiedau} to enable access of P\&ID data over OPC UA communication platforms. Also, In \cite{Holm}, ISO 15926 and IEC 62424, i.e. two different standards for computer-accessible structural model descriptions, which have been conceived for the modeling of process plants, have been compared.

\section{Case study}
\begin{figure}[htbp]
\centerline{\includegraphics[width=0.45\textwidth]{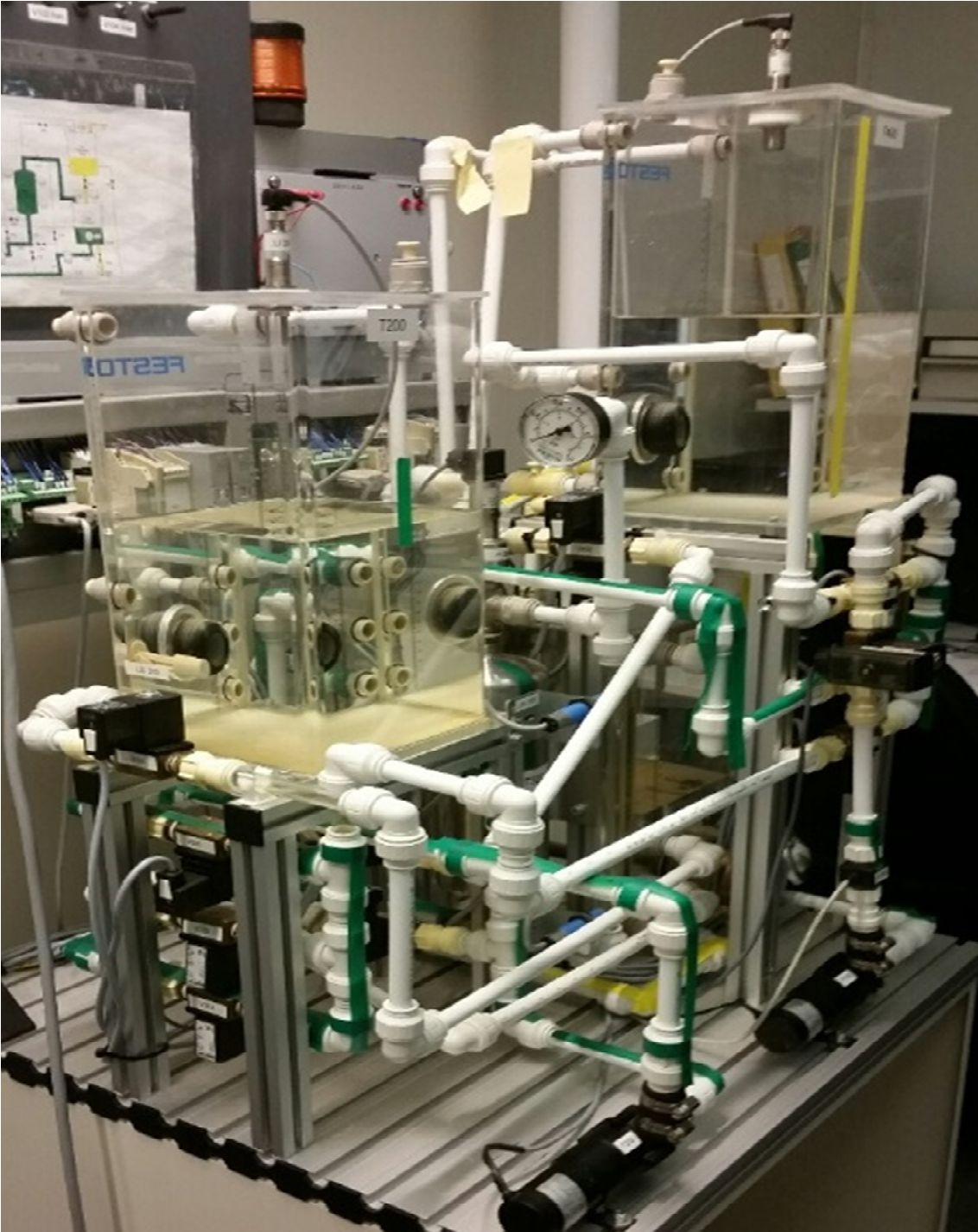}}
\caption{Case process}
\label{Case process}
\end{figure}

The case study is a thermo-hydraulic water process (Fig.~\ref{Case process}). The functionality of the process is not important for the aims of this article, but interested readers will find more details in \cite{Martinez1,Martinez2,Martinez3,Sierla}.

\begin{figure}[htbp]
\centerline{\includegraphics[width=0.45\textwidth]{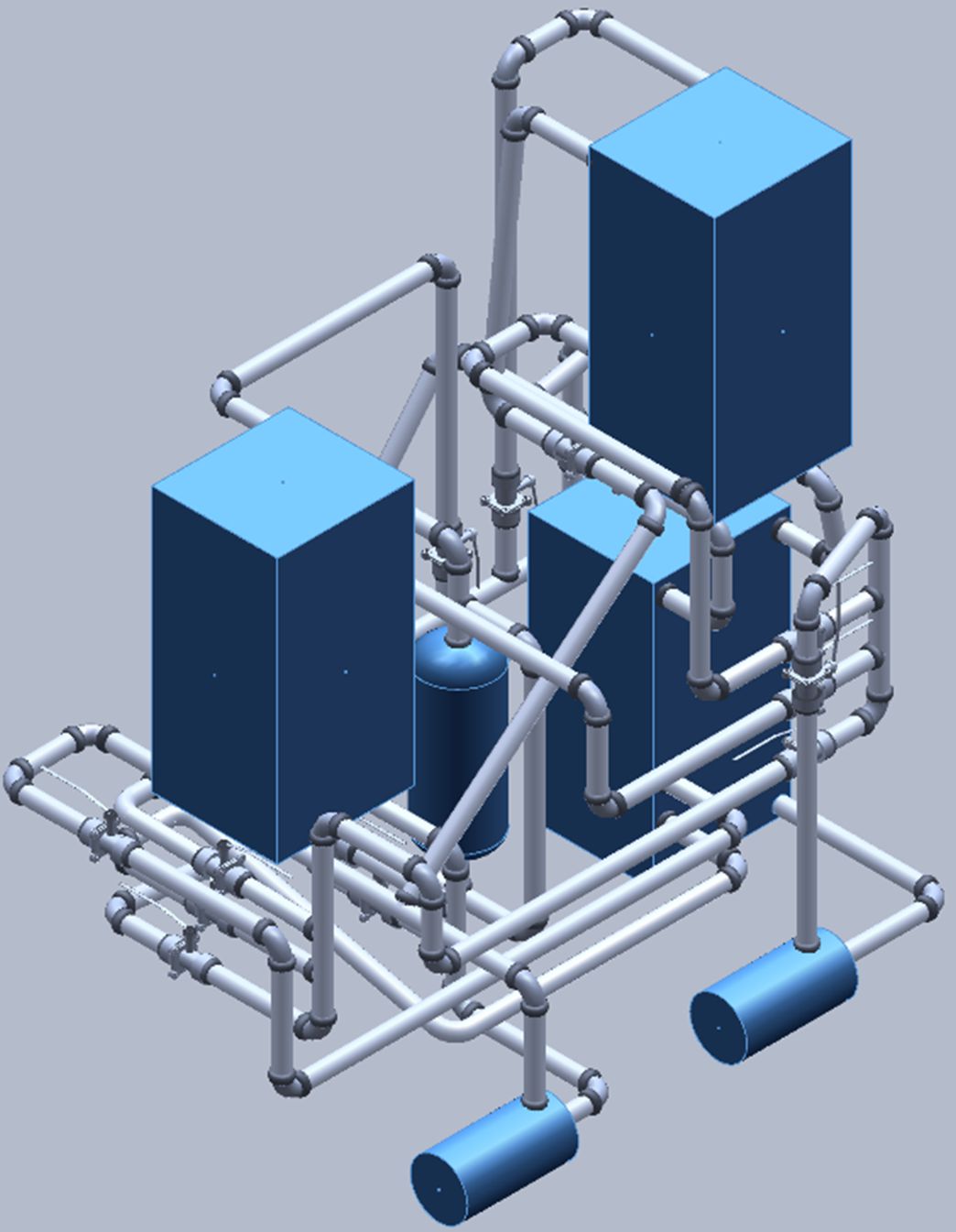}}
\caption{3D CAD model of the plant in Fig.~\ref{Case process}.}
\label{3D_CAD}
\end{figure}

The process has been modelled in the Intergraph Smart 3D tool (Fig.~\ref{3D_CAD}), which is capable of exporting PCF files. The model includes 10 pipelines, each of which has its corresponding PCF file. Pipelines may have branches. The endpoint of a pipeline is either a nozzle of process equipment or an open endpoint, referencing another open end point in another PCF file.

\begin{figure*}[htbp]
\centerline{\includegraphics[width=0.80\textwidth]{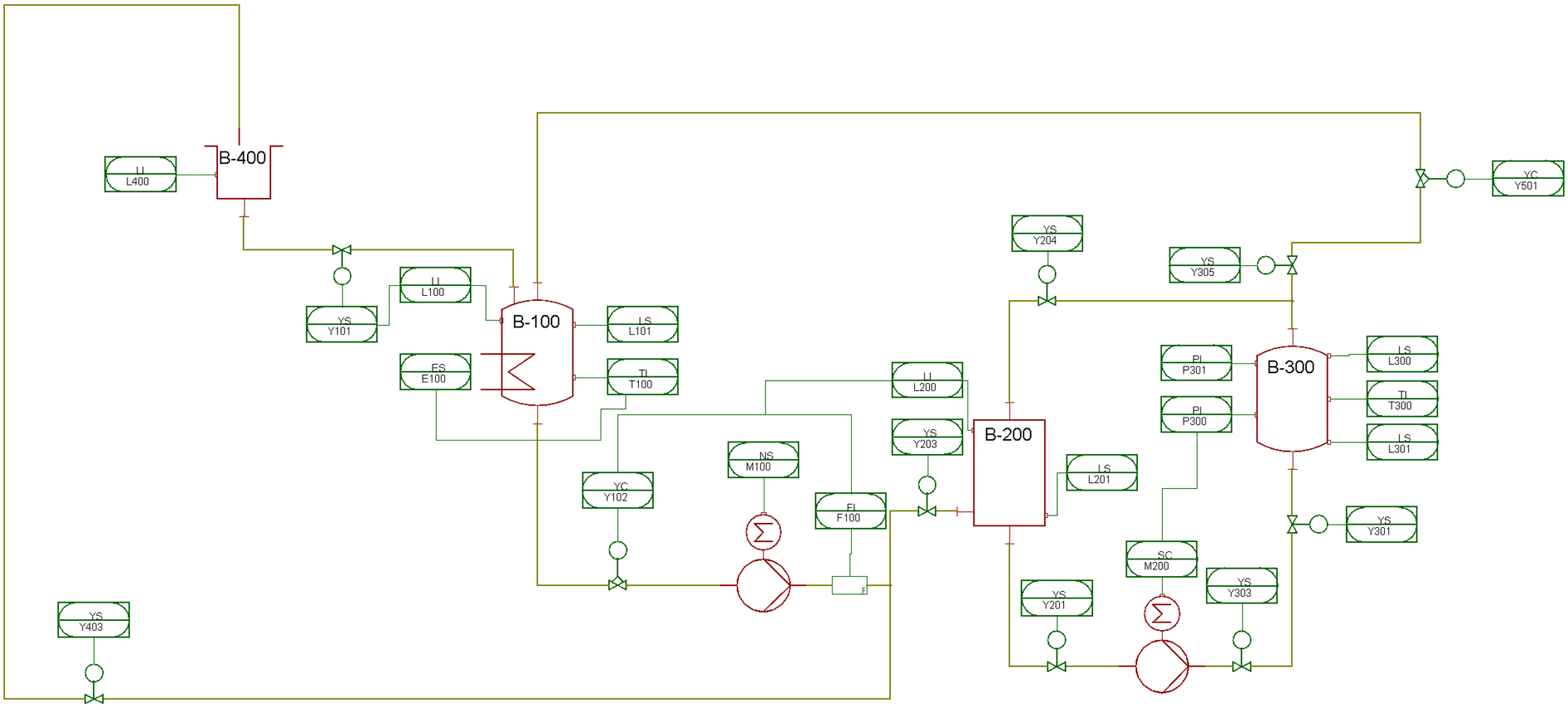}}
\caption{P\&ID of the plant in Fig.~\ref{Case process}.}
\label{pid}
\end{figure*}

A P\&ID has been developed in the SmartPlant P\&ID and exported with its ISO 15926 export tool. The exported file conforms to the Proteus 3.6 XML schema.  Fig.~\ref{pid} presents a visualization of the exported XML file. It is notable that Fig.~\ref{pid} includes only the main pipelines, while Fig.~\ref{3D_CAD} includes all of the pipelines. In general, such differences may be encountered in industrial plants, especially when working with design documents originating from different phases of the plant life-cycle.

\section{Methodology}

Directed graphs with node labels are chosen as the abstraction level for step 2 of the procedure presented in Section I, as it is anticipated that they can support the matching activity in step 3. Such graphs have been previously successfully applied to matching industrial process plant design \cite{Rantala}. Matching of P\&IDs and 3D models has not yet been attempted. The research goal stated in Section I can thus be elaborated as follows: to generate directed graphs with relevant node labels from P\&IDs in Proteus DEXPI format and 3D CAD models in PFC format. Since the goal is to raise the level of abstraction, it is intended that the graph capture only a part of the information in the source document. The ideal level of information to be captured depends on the needs of steps 3 and 4 of the procedure introduced in Section I, so this is a discussion that is initiated in this paper and continued in further research. However, previous research on graph matching has shown that performance has been improved by graph simplification methods that have discarded details related to piping \cite{Rantala}, so our starting point in this paper is that more detail is not necessarily better. The graph is specified as a set of node \(N\) and a set of directed edges \(E\). Each edge is specified by source and target nodes \(e_{source}\) and \(e_{target}\), which are elements of \(N\).

\subsection{A Generating a graph from a Proteus XML file}  \label{Generationg graph by XML}
\begin{figure}[htbp]
\centerline{\includegraphics[width=0.5\textwidth]{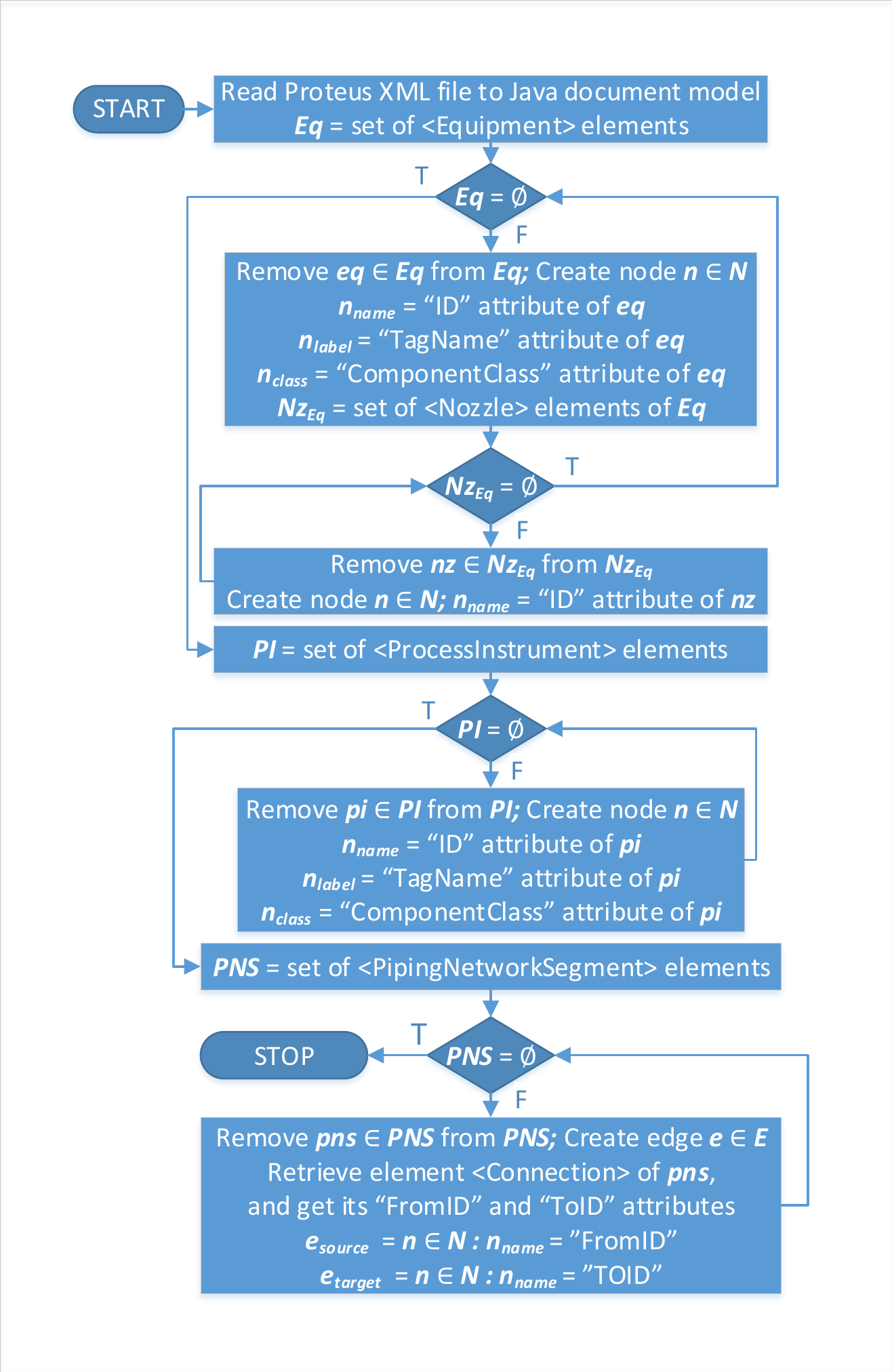}}
\caption{Flowchart for generating a directed graph from an XML file conforming to Proteus XML schema.}
\label{xml}
\end{figure}

Fig.~\ref{xml} presents a flowchart for generating the graph from an XML file conforming to the Proteus XML schema. The procedure extracts the connections between elements of the physical process or the control system, as opposed to graphical connections in the diagram.  The \(<Connection>\) element of \(<PipingNetworkSegment>\) elements specifies connections nozzles of tanks or pumps to each other. In some cases, a \(<PipingNetworkSegment>\) connects to a valve and in some cases the valve is skipped, in the sense that the piping network segments have no information to specify that a valve was along that segment. Whether this occurs depends on the way in which the engineer uses the P\&ID tool. For the control system, connectivity is specified in terms of the <Connect> elements of \(<SignalLine>\) elements. However, it was discovered that these connections are between two elements of type \(<InstrumentComponent>\), which in our case are valves, heating elements, pump motors or generic actuators of unspecified type. Thus, this would result in many small stand-alone graphs not connected to the graph generated from \(<PipingNetworkSegments>\). For example there is a \(<SignalLine>\) between the temperature sensor TI-T100 to the heating element ES-E100 (see tank B-100 in Fig.~\ref{pid}), but ES-E100 is not logically connected to the tank; it is just drawn next to the tank so that a human will understand that it refers to the heating element in the tank. Thus, although extraction of \(<InstrumentComponent>\) and \(<SignalLine>\) elements was implemented, it was concluded after examining the results for the case study that it is very questionable whether these would add value to the generated graph. Thus, the procedure in Fig.~\ref{xml} does not examine these elements. It is understood that further research is needed to determine the ideal level of detail for graphs generated from Proteus XML.

The flowchart in Fig.~\ref{xml} attaches 3 kinds of labels to nodes: \(n_{name}\) (a unique id), \(n_{lable}\) (a tag for human readable presentation) and \(n_{class}\) (which specifies the type of component and may be used later for graph matching purposes).

\subsection{Generating a graph from a PCF file}

\begin{figure*}[htbp]
\centerline{\includegraphics[width=0.76\textwidth]{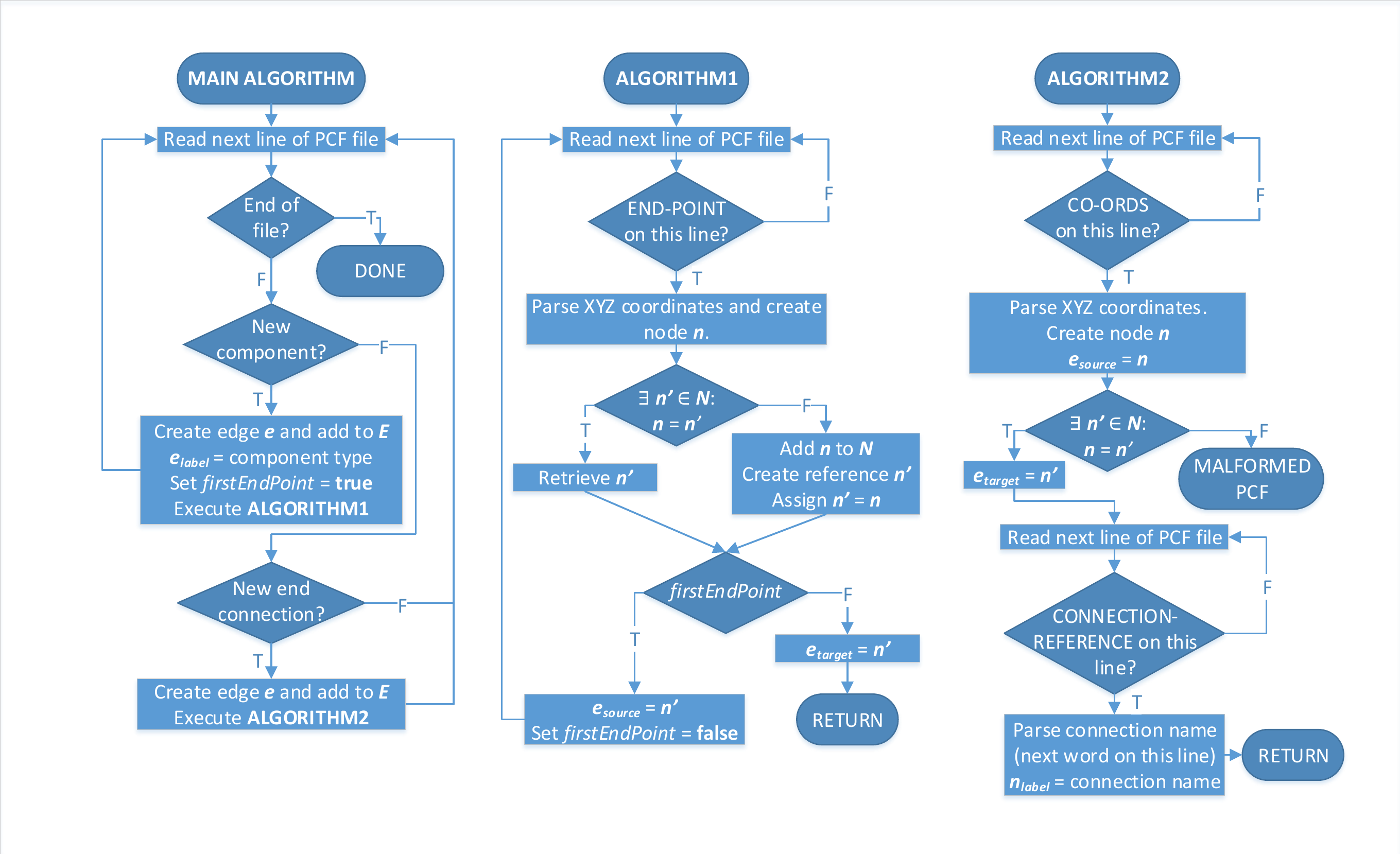}}
\caption{Flowchart for generating a directed graph from a PCF file.}
\label{pcf}
\end{figure*}

Fig.~\ref{pcf} presents a procedure for generating a directed graph from a PCF file. The ‘New Component?’ element of the MAIN ALGORITHM examines components of type PIPE, WELD and VALVE. The PCF also defines TEE-STUB elements, but the branches in the pipelines can be captured in the graph without examining these elements. Each component has two END-POINT lines in the PCF file, which specify 3D coordinates. Each such coordinate will result in a node in the graph. The two END-POINTs are used to define an edge between the nodes that they define. The edge is labelled with the type of component; the types relevant for our case study are ‘Pipe’, ‘Weld’ and ‘Valve’. Thus, these nodes do not correspond to nodes in the graph generated from the P\&ID. ALGORITHM2 in Fig.~\ref{pcf} extracts end connections from the PCF and generates nodes for them as well. The end connection of a pipeline is either a nozzle of process equipment or an open endpoint, referencing another open endpoint in another PCF file. In the case of a nozzle of process equipment, the created node will have a direct correspondence to a node generated from the P\&ID. The label of this node generated by ALGORITHM2 is a string that combines tag and component type information, thus merging information similar to \(n_{lable}\) and \(n_{class}\) generated by the algorithm Fig.~\ref{xml}.

\section{Result}
\subsection{Graph abstraction of the 2D model}
\begin{figure}[htbp]
\centerline{\includegraphics[width=0.5\textwidth]{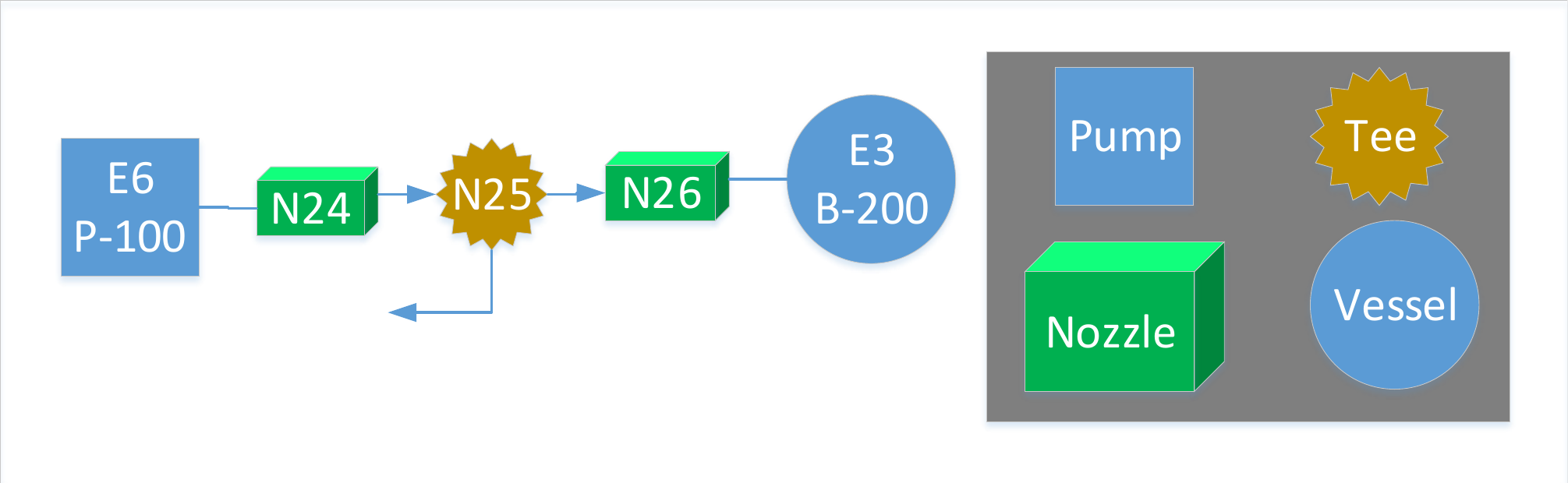}}
\caption{Excerpt of generated graph.}
\label{r1}
\end{figure}

\begin{figure}[htbp]
\centerline{\includegraphics[width=0.30\textwidth]{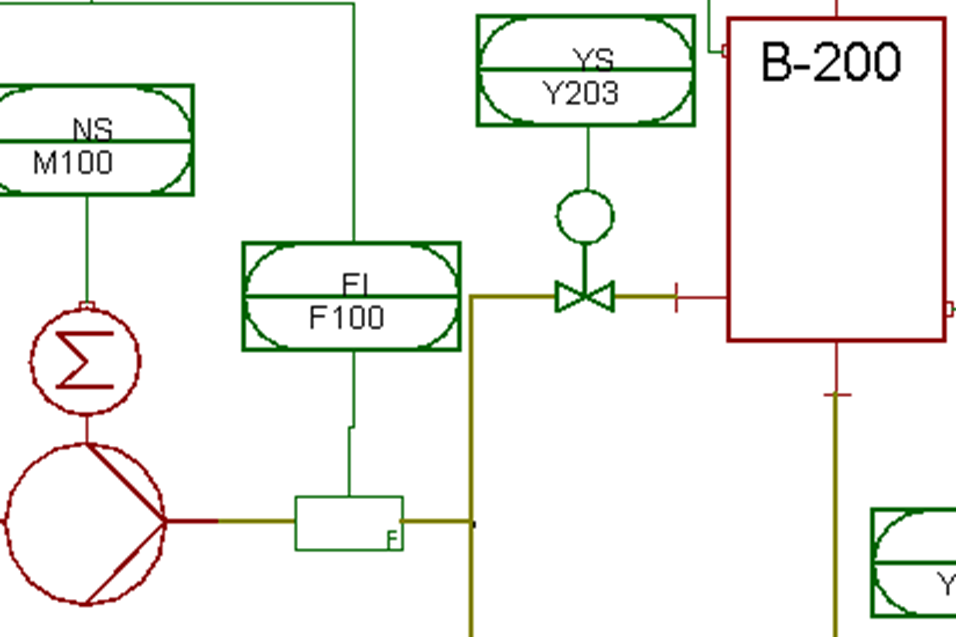}}
\caption{Excerpt of P\&ID corresponding to the graph in Fig.~\ref{r1}.}
\label{r2}
\end{figure}

The algorithm in Fig.~\ref{xml} was implemented in Java and applied to a Proteus XML file corresponding to the P\&ID in Fig.~\ref{pid}. Fig.~\ref{r1} shows an excerpt of the generated graph, which was drawn manually from the list of nodes and edges exported by the algorithm. Fig.~\ref{r2} shows the corresponding part of the P\&ID. It is notable that in this case, the graph generation skips the inline instruments, the flow meter and the valve, for reasons explained in \ref{Generationg graph by XML}.

\begin{figure}[htbp]
\centerline{\includegraphics[width=0.5\textwidth]{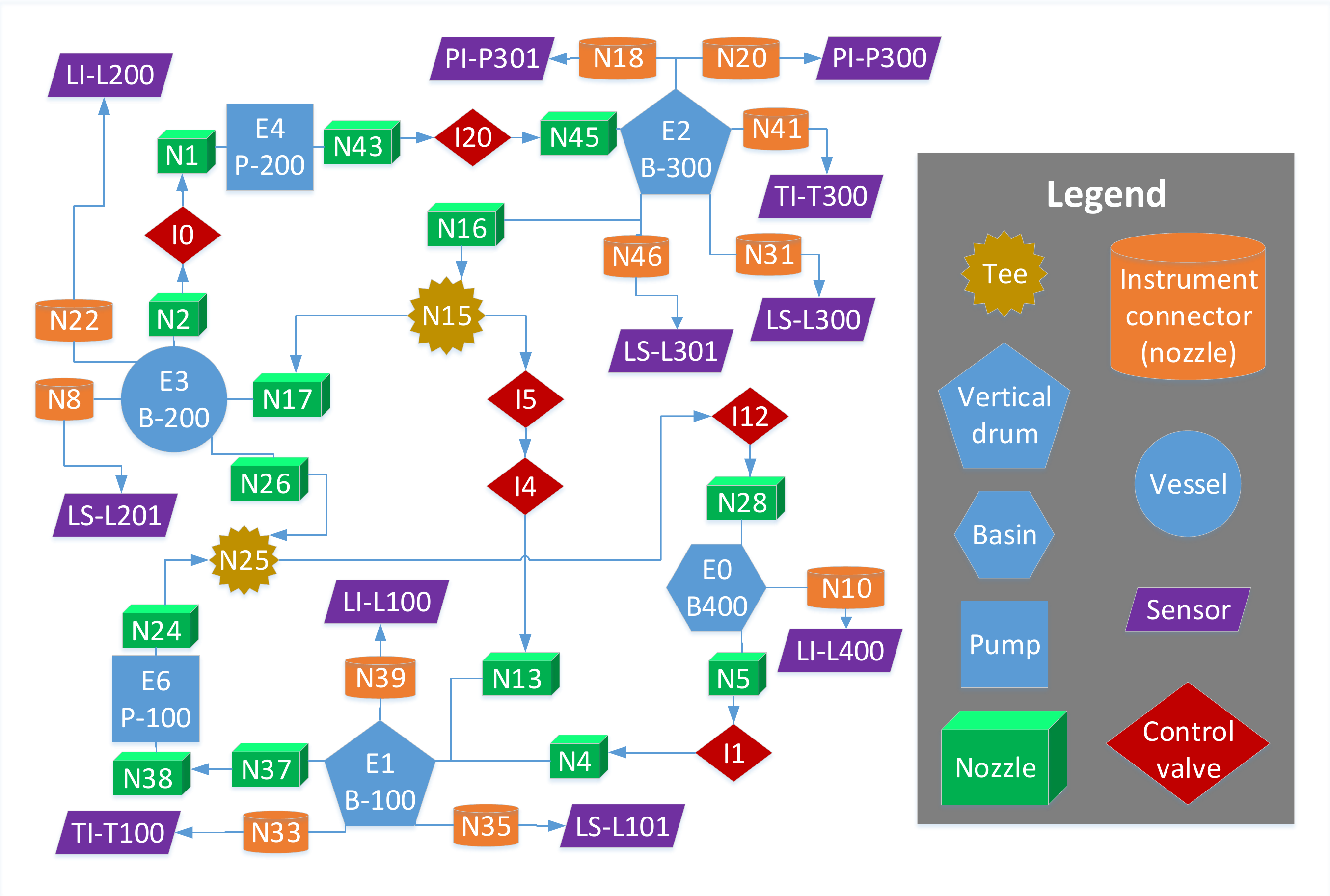}}
\caption{Complete graph generated from P\&ID.}
\label{Complete_Graph}
\end{figure}

Fig.~\ref{Complete_Graph} shows the complete generated graph, which was drawn manually from the list of nodes and edges outputted by the algorithm. Referring to the node labels created by the algorithm in Fig.~\ref{xml}, \(n_{name}\) is a very long and hardly human readable id, which has been replaced in our implementation by a unique short human readable id such as E3, N16, or I5. \(n_{lable}\) is a tag for human readable presentation and may not always be present in the XML file; in Fig.~\ref{Complete_Graph} it has been used in addition to the id (in case of tanks and pumps) and instead of the id in case of sensors. \(n_{class}\) provides the information on component types in the legend.

\subsection{Graph abstraction of the 3D model}

\begin{figure}[htbp]
\centerline{\includegraphics[width=0.5\textwidth]{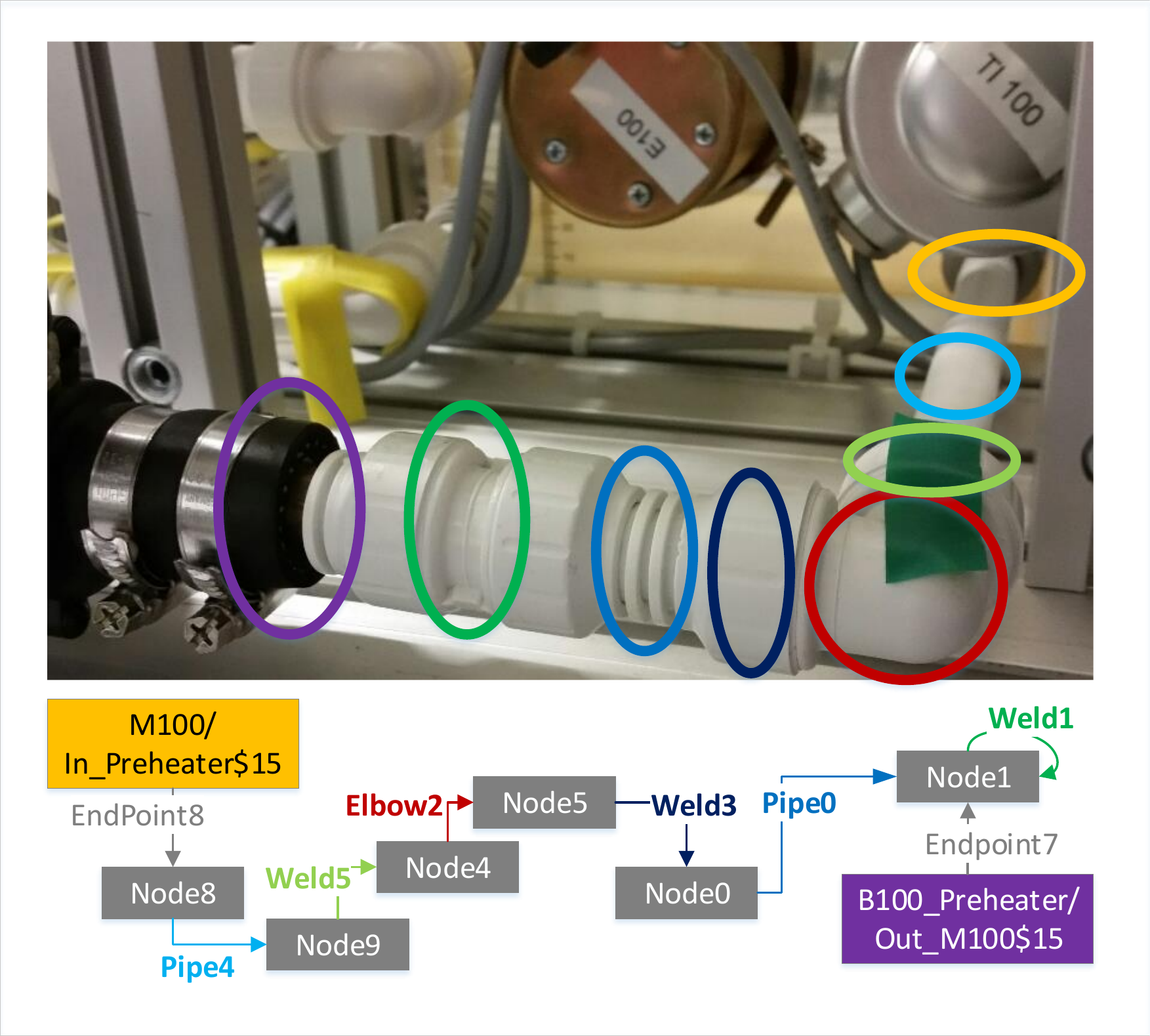}}
\caption{Graph for the pipeline from tank 100 to pump 100}
\label{tank100}
\end{figure}

Fig.~\ref{tank100} shows the graph generated from the PCF of the simplest of the 10 pipelines in Fig.~\ref{3D_CAD}. The graph was drawn manually from the list of nodes and edges outputted by the algorithm in Fig.~\ref{pcf}. Color coding is used to show the corresponding process components on the photo of the pipeline.

\begin{figure}[htbp]
\centerline{\includegraphics[width=0.5\textwidth]{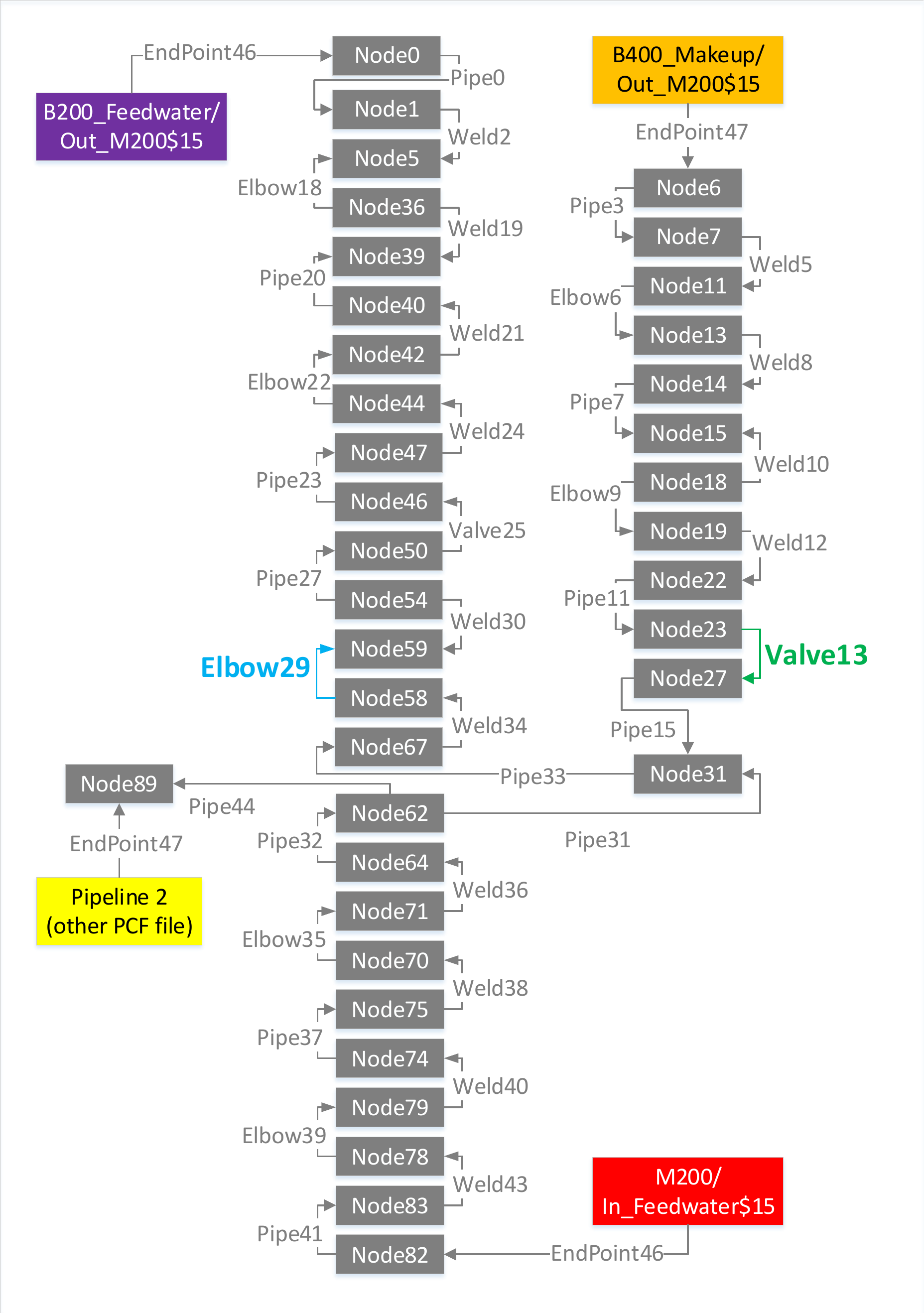}}
\caption{Graph generated from the PCF of the pipeline from tank 200 to pump 200.}
\label{tank200}
\end{figure}

\begin{figure}[htbp]
\centerline{\includegraphics[width=0.45\textwidth]{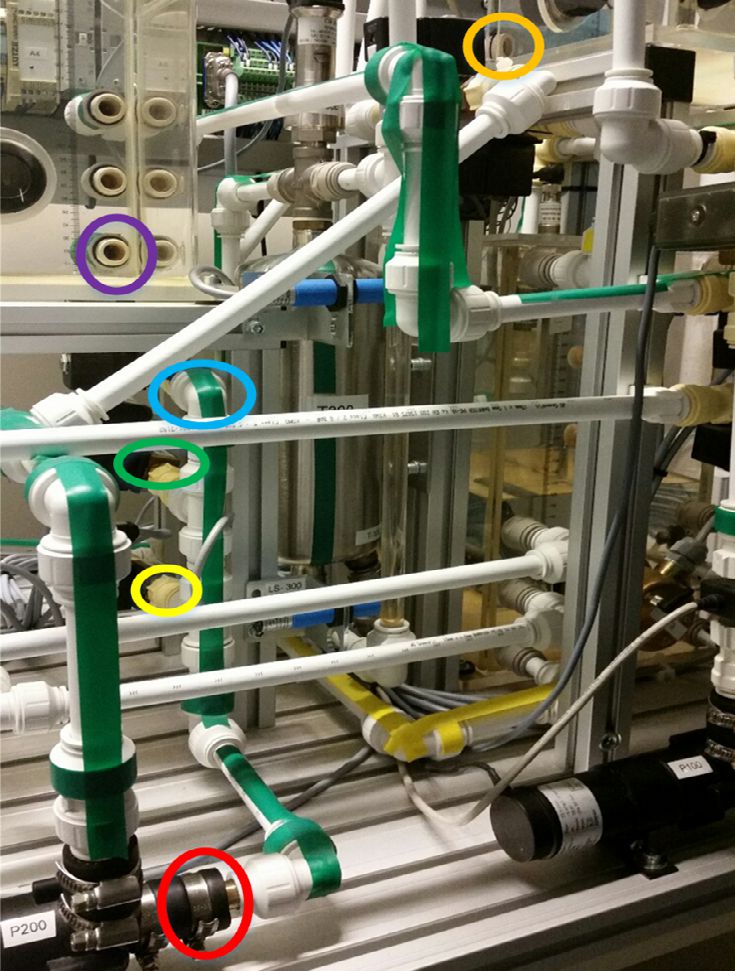}}
\caption{Marking the color coded nodes and edges in Fig.~\ref{tank200} with colored ellipses on the relevant process components on the photo of the pipeline.}
\label{tank200_1}
\end{figure}

Fig.~\ref{tank200} repeats the experiment on a more complex pipeline with branches and valves, and Fig.~\ref{tank200_1} uses color coding to mark the corresponding elements on a photo of the pipeline.

It is notable that the edge directions do not correspond to flow directions, which may be a major issue for matching approaches based on directed graphs. Three possible solutions are proposed for further research:
\begin{enumerate}
\item	It would be possible to examine the FLOW attributes of the PCF. However, these are optional attributes and it cannot be assumed that modelers define them, so this approach is not recommended.
\item	The START-CO-ORDS attribute of the pipeline can be used. This can be used to identify the end connection node on the pipeline from which the flow originates. The solution depends on assuming that in 1 PCF file there are branches but no loops and that flows in all branches are away from the node at START-CO-ORDS. START-CO-ORDS is an optional attribute, but this assumption could be enforced by a semi-automatic solution that asks the users to specify the start node for each pipeline. If there is a usable interface which allows the user to select from options in a drop-down menu, the manual workload would be minimal. In this case, the edge directions can be fixed to correspond to flow directions by treating the graph generated from the PCF file as a tree with the node at START-CO-ORDS as the root node. The graph could then be processed with a tree traversal algorithm, so that edge directions are fixed to always point away from the root.
\item	The ingoing and outgoing flows at pumps are specified in the end connection information of the PCF. In case of pipelines without pumps, in which the flow is caused by gravity, the elevations of the endpoints can be used to infer the flow direction. This could be used to overcome the need for manual input in solution 2 in case START-CO-ORDS has not been used.
\end{enumerate}

\section{Discussion}

\begin{figure}[htbp]
\centerline{\includegraphics[width=0.5\textwidth]{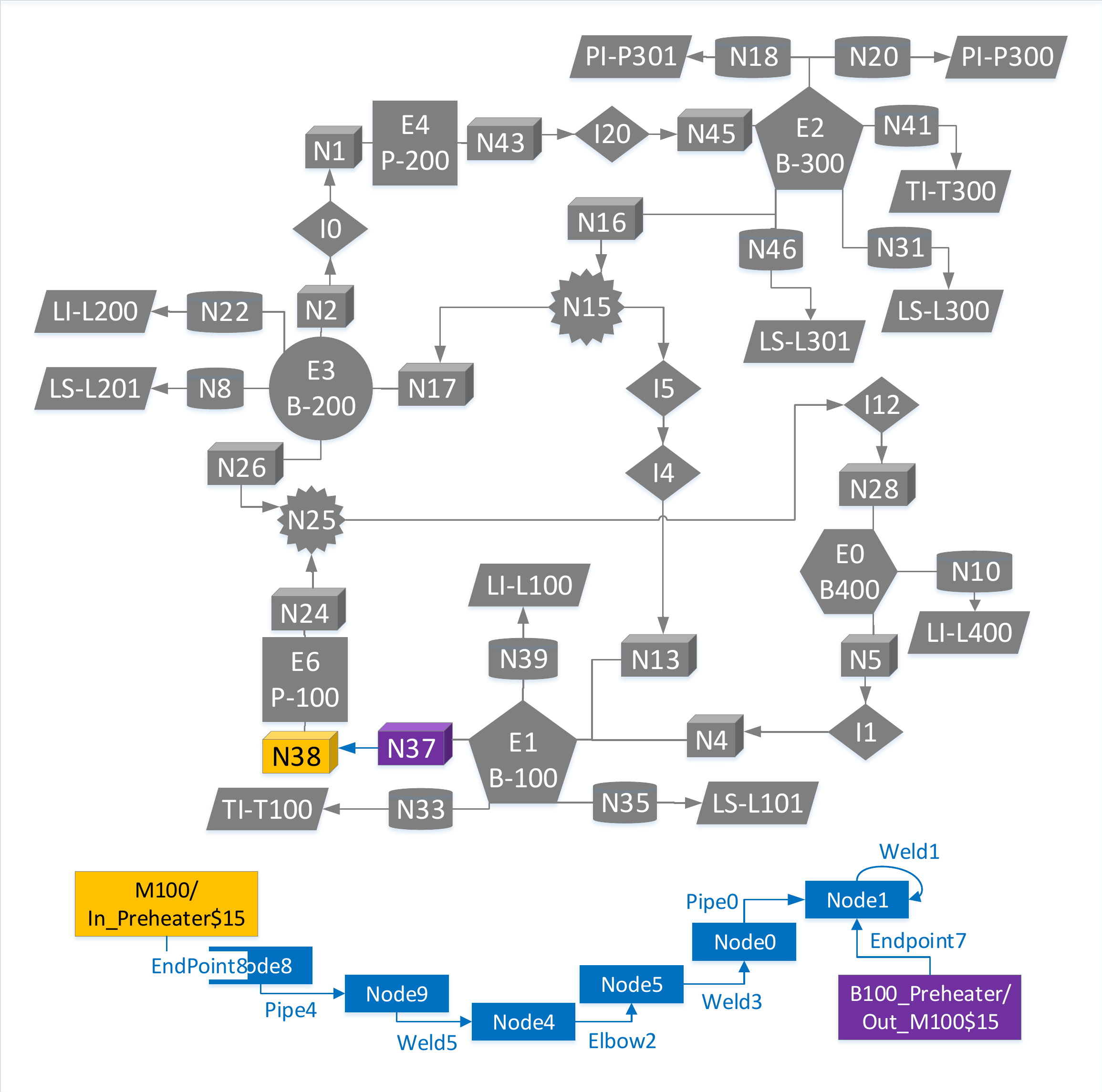}}
\caption{Matching the graph in Fig.~\ref{tank100} to the graph in Fig.~\ref{Complete_Graph}.}
\label{matching1}
\end{figure}

Fig.~\ref{matching1} matches the PCF generated graph in Fig.~\ref{tank100} to the Proteus generated graph in Fig.~\ref{Complete_Graph}. Color-coding is used to show the matching elements. The blue color makes it clear how the graphs are at a different level of abstraction. Graph simplification methods such as presented in \cite{Rantala} could readily be applied to eliminate this difference; however, the raw outputs are presented in Fig.~\ref{matching1}, since the ideal graph simplification approach is a matter of further research.

\begin{figure}[htbp]
\centerline{\includegraphics[width=0.5\textwidth]{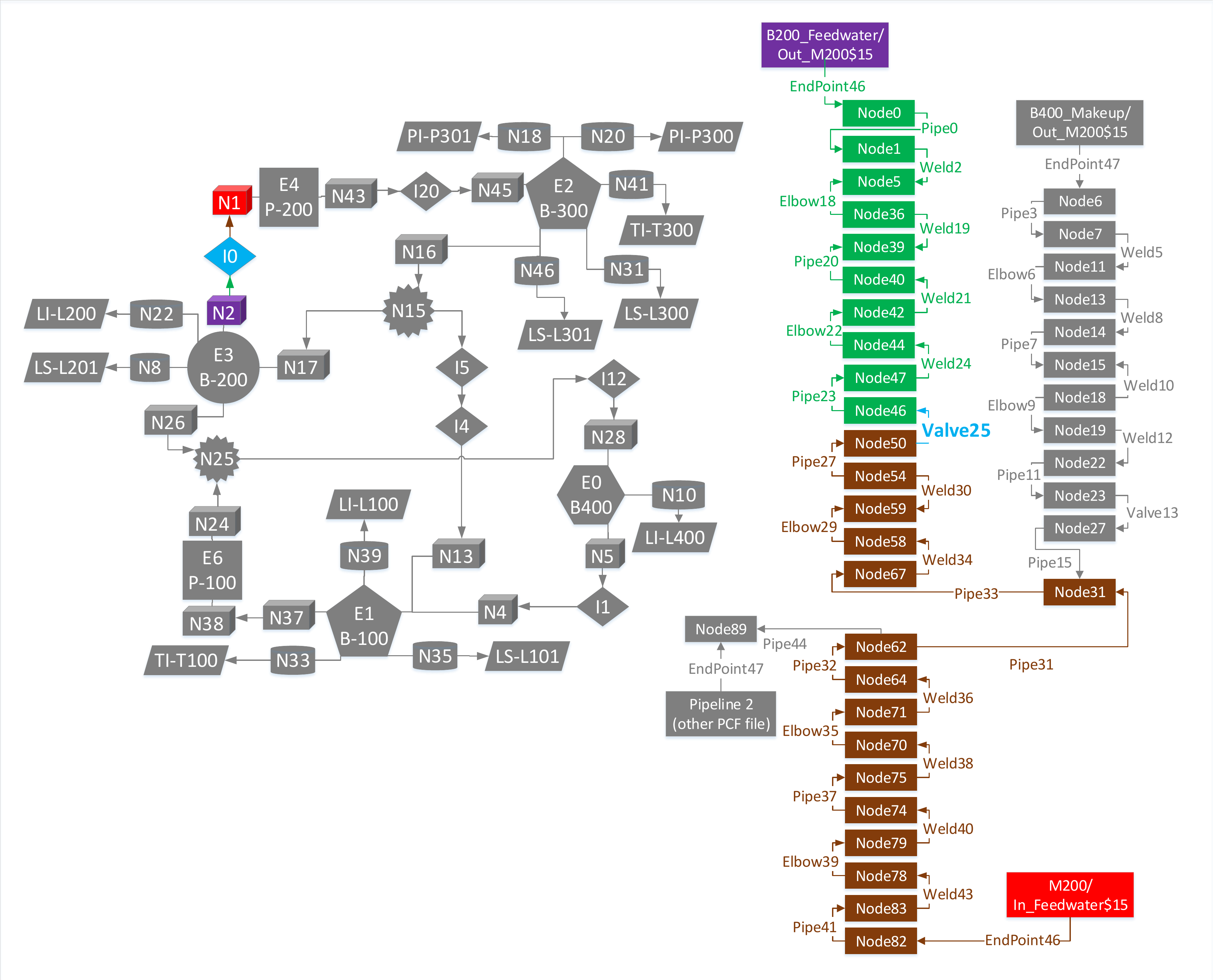}}
\caption{Matching the graph in Fig.~\ref{tank200} to the graph in Fig.~\ref{Complete_Graph}.}
\label{matching2}
\end{figure}

Fig.~\ref{matching2} matches the PCF generated graph Fig.~\ref{tank200} to the Proteus generated graph in Fig.~\ref{Complete_Graph}. It is notable that some parts of the PCF generated graph could not be matched, as they correspond to pipelines not included in the simplified P\&ID. As discussed in Section III, such a scenario is likely to occur in industrial practice over the plant lifecycle and solutions developed in further work should be robust against these scenarios.

The color-coding in Fig.~\ref{matching1} and Fig.~\ref{matching2} was added manually. The automation of this matching work belongs to step 3 of the procedure introduced in Section I and is expected to be done in further work by graph matching techniques similar to \cite{Rantala}.
It is notable that the graphs generated by the algorithms in Fig.~\ref{xml} and Fig.~\ref{pcf} are a straightforward abstraction of the information in the source formats. Thus, they may not be ideal inputs for graph matching methods in further work. In particular, nodes in the graph generated from a P\&ID correspond to process components and have a label \(n_{class}\), which specifies the type of component. However, the PFC file specifies components such as pipe segments, welds and valves with result in edges. In other words, nodes in the P\&ID graph may correspond to edges in the PCF graph (such as the indigo coded elements in Fig.~\ref{matching2}).

\section{Conclusion}
To summarize the discussion, a preprocessing phase may be needed before graph matching to address the identified disparities between the graphs generated from the 2D and 3D sources. In particular, piping simplifications algorithms as in \cite{Rantala} could be applied to the graphs generated from the 3D CAD to arrive at the same level of details as in the P\&ID graphs. Additional novel preprocessing algorithms are required to address disparities such as valves being represented as nodes in the 2D graph and as edges in the 3D graph. Finally, the findings suggest that level of tool support and industry standardization for capturing flow directions may be insufficient for the development of robust and general solutions for generating directed graphs from 3D CAD models. In this case, one viable option is to work with undirected graphs, since according to previous research the direction information is only used to variants of the graph matching algorithm, such as the ‘anchor similarity measure’ in \cite{Rantala}. After these preprocessing steps, it is reasonable to expect the graph matching will give good results, since the graphs to be matched have similar structure and level of detail. The matching will provide the basis for integrating the 2D and 3D information to a single digital plant model.

\section{Acknowledgements}
This work was partially supported by Business Finland project SEED (grant 4153/31/2019.)

\end{document}